\documentclass[aps,twocolumn,showpacs,pra,floatfix,preprintnumbers]{revtex4}

\usepackage{amsmath}
\usepackage{amssymb}
\usepackage{pifont}
\usepackage{graphicx}

\usepackage{hhline}

\usepackage{graphicx}

\newcommand{\ket}[1]{\left\vert#1\right\rangle}
\newcommand{\beq}{\begin{equation}}
\newcommand{\eeq}{\end{equation}}
\newcommand{\bea}{\begin{eqnarray}}
\newcommand{\eea}{\end{eqnarray}}

\makeatletter
\def\btt#1{\texttt{\@backslashchar#1}}
\DeclareRobustCommand\bblash{\btt{\@backslashchar}}
\makeatother
\def\Bid{{\mathchoice {\rm {1\mskip-4.5mu l}} {\rm
{1\mskip-4.5mu l}} {\rm {1\mskip-3.8mu l}} {\rm {1\mskip-4.3mu l}}}}

\begin{document}

\title{Heralding an Arbitrary Decoherence-Free Qubit State} 
\author{C. Allen Bishop\footnote{Email: bishopca@ornl.gov}, Ryan S. Bennink, Travis S. Humble, Philip G. Evans}
\affiliation{$^{1}$ Quantum Information Science Group, Computational Sciences and Engineering Division, 
Oak Ridge National Laboratory, Oak Ridge, Tennessee 37831-6418, USA}
\author{Mark S. Byrd} 
\affiliation{ $^{2}$ Department of Physics, Southern Illinois University, 
Carbondale, Illinois 62901-4401, USA}

\date{\today}

\begin{abstract}
We present an optical device which is capable of 
heralding a variety of 
DFS states which protect against collective noise. Specifically, it can 
prepare all three basis states which span a DFS qutrit as well as an arbitrarily 
encoded DFS qubit state. We also discuss an interferometric technique for determining the amplitudes 
associated with an arbitrary encoding. The heralded state may find use in coherent optical systems which exhibit 
collective correlations.

\end{abstract}


\maketitle



\emph{Introduction.---}
Error-avoiding quantum codes offer a passive approach to the protection of quantum information \cite{Zanardi}. (For reviews of this subject, see \cite{Lidar1,Byrd,Lidar}.) Decoherence-free subspaces, or more generally, decoherence-free/noiseless subsystems (DFSs/NSs), provide a promising avenue for protection when a symmetry in the system-environment interaction decouples the NS from the 
environment. Methods for identifying 
noiseless subsystems have been presented \cite{Kribs, Knill,Mahler, Wang} and the predicted immunity to specific noise processes have been verified experimentally in various settings. In particular, experiments using trapped ions \cite{Kielpinski,Roos,Langer,Monz}, nuclear magnetic resonance (NMR) systems \cite{Viola,Fortunato,Ollerenshaw}, and photonic systems \cite{Kwiat1,Mohseni,Bourennane} have shown the benefits of using  decoherence-free subspace or subsystem encodings to limit the effects of decoherence.

The first experimental investigation of a decoherence-free subspace was performed and reported by Kwiat {\it{et al.}} \cite{Kwiat1}. Using linear optics, parametric down-conversion (PDC), and postselection, they 
were able to demonstrate the resilience of the singlet state $\ket{\psi^-} = (\ket{VH} - \ket{HV})/\sqrt{2}$ to engineered collective-dephasing channels thereby establishing an excellent agreement between experiment and theory. Later, Mohseni {\it{et al.}}, using 
optical rail qubits \cite{Mohseni}, 
and Ollerenshaw {\it{et al.}}, using NMR \cite{Ollerenshaw}, independently
provided the first experimental demonstrations of how decoherence-free subspaces could be used to improve the performance of quantum algorithms. In both of these experiments, noise was induced in a non-collective way by the authors. During that same year, Eibl {\it{et al.}} recognized that a particular state which is naturally emitted during a double-pair PDC emission process \cite{Weinfurter} remains invariant under the collective interactions \cite{Eibl} (see footnote therein). This state 
was first introduced by Kempe {\it{et al.}} \cite{Kempe} as a four-qubit collective decoherence free subspace. Both of the singlet states which span the four-qubit DF subspace were postselected in Ref.~\cite{Bourennane} along with a particular 
superposition of the two. Later, a proposal was made for the postselection of some, but not all, of the superposition states spanned by the two singlets of the four-qubit DF subspace \cite{Zou}. This effort was extended by Gong {\it{et al.}} with a proposal for an optical device which is capable of preparing, through a postselection strategy, an arbitrary four-qubit DF subspace qubit state \cite{Gong}. However, polarization-insensitive beam-splitters of variable reflectivity are required for use as the singlet state amplitudes are functionally dependent on them. In practice, changing the reflectivity of a beam-splitter amounts to replacing the beam-splitter altogether. This feature limits the switching rate between different qubit states. Furthermore, postselected states in general suffer from several disadvantages. The most significant drawback comes from the fact that state verification inherently destroys the state which was selected. This limits the extent of utilization in many practical instances. For example, a postselected state 
may be used to transmit information from one point to another, however, the sender has no way of knowing when the message state was transferred until the receiver actually measures it. Postselection also limits the degree to which one can scrutinize a quantum state since the measurement process typically becomes difficult in more than one basis.

Here, we present a proposal for an optical device capable of heralding an arbitrary decoherence-free qubit state encoded into the triply-degenerate four-qubit DFS. Successful state preparation is heralded by coincident detection of two auxiliary photons. Rotation over the entire Bloch sphere can be achieved using a single phase shifter and a single polarization rotator. By including two additional wave-plates, the device can further prepare all three basis states of a DFS qutrit.  
This encoding protects the data qubit from arbitrary collective noise effects; collective rotations, collective phase drifts, as well as combinations of both types. In order to read out these encodings, we provide a method for distinguishing each of the three DFS qutrit basis states in the logical basis. 

We will begin with a discussion of the mathematical structure inherent to the four-qubit DFS. We emphasize that we are encoding into a subsystem rather than the four-qubit DF subspace mentioned above. The details concerning the DFS state generator as well as the decoding interferometer will then be provided. We will use the terms noiseless subsystem and decoherence-free subsystem interchangeably throughout the text.

\emph{The logical states.---}
\label{sec:encoding}
Using Young's tableau \cite{Cornwell,MByrd}, we find the following decomposition of the state space of four physical qubits: 
${\bf{2}}\otimes{\bf{2}}\otimes{\bf{2}}\otimes{\bf{2}}= 
{\bf{1}}\oplus{\bf{1}}\oplus{\bf{3}}\oplus{\bf{3}}\oplus{\bf{3}}\oplus{\bf{5}}$. Here ${\bf{N}}$ denotes an $N$-dimensional irreducible representation of ${\text{SU}}(2)$. The singlet states arising in this decomposition were considered in the works 
of Refs~\cite{Bourennane,Zou,Gong}. We will instead consider the triply-degenerate ${\bf{3}}$ representations and use four physical qubits to support an encoded collective-DFS qutrit $\ket{\Psi_{DFS}}=\nu_0\ket{0_L}+\nu_1\ket{1_L}+\nu_2\ket{2_L}$,
with $|\nu_0|^2+|\nu_1|^2+|\nu_2|^2=1$. Each logical basis state can be expanded as 
\bea
\label{eq:qs2}
\ket{0_L} &=& \omega_{0,1}\ket{0_L^1} + \omega_{0,2}\ket{0_L^2} + 
\omega_{0,3}\ket{0_L^3}, \\ 
\ket{1_L} &=& \omega_{1,1}\ket{1_L^1} + \omega_{1,2}\ket{1_L^2} + 
\omega_{1,3}\ket{1_L^3}, \\
\ket{2_L} &=& \omega_{2,1}\ket{2_L^1} + \omega_{2,2}\ket{2_L^2} + 
\omega_{2,3}\ket{2_L^3}, 
\eea
where $\sum_{k=1}^3|\omega_{i,k}|^2 = 1 
\;\; (i=0,1,2)$. The physical qubits we consider are photons, each of which 
being described in the polarization basis $\{ \ket{H}, \ket{V} \}$. The eigenstates $\ket{j,m}$ of the angular momentum operator $J^2$ can be calculated using standard Clebsch-Gordan algebra. For the logical zero region we find 
\bea
\label{eq:zero}
\ket{0_L^1} &=& (\ket{\psi^+}\otimes \ket{VV}-\ket{VV} \otimes 
\ket{\psi^+})/\sqrt{2,} \nonumber \\
\ket{0_L^2} &=& (\ket{HHVV} - \ket{VVHH})/\sqrt{2}, \nonumber \\
\ket{0_L^3} &=& (\ket{HH} \otimes \ket{\psi^+}-\ket{\psi^+} \otimes 
\ket{HH})/\sqrt{2}, 
\eea
with $\ket{\psi^\pm} = (\ket{VH} \pm \ket{HV})/\sqrt{2}$. (We will also refer to the states $\ket{\phi^\pm}= (\ket{HH} \pm \ket{VV})/\sqrt{2}$ later.) Those which 
span the logical one and two states are given by
\bea
\label{eq:onetwo}
\ket{1_L^1} &=& \ket{VV} \otimes \ket{\psi^-}, \;\;
\ket{2_L^1} = \ket{\psi^-} \otimes \ket{VV}, \nonumber \\
\ket{1_L^2} &=& \ket{\psi^+} \otimes \ket{\psi^-}, \;\:
\ket{2_L^2} = \ket{\psi^-} \otimes \ket{\psi^+}, \nonumber \\
\ket{1_L^3} &=& \ket{HH} \otimes \ket{\psi^-}, \;
\ket{2_L^3} = \ket{\psi^-}\otimes \ket{HH}. 
\eea

There is a great deal of freedom in the DFS encoding process. The only requirement is that
$\sum_{k=1}^3 |\omega_{i,k}|^2 = 1$ for $i=0,1,2.$ For the moment, we will restrict our attention 
to the space spanned by $\{ \ket{1_L},\ket{2_L} \}$ and present a method for heralding an 
arbitrary DFS qubit state $\cos{\theta} \ket{1_L} + \sin{\theta}e^{i\phi} \ket{2_L}$. We emphasize here that this 
paper discusses a method for heralding an arbitrary DFS qubit state as well as all three basis states which span a 
DFS qutrit. We have not found a way to herald an arbitrary DFS qutrit state. That being said, we find it 
convenient to initialize the noiseless subsystem using the states $\ket{1_L^2} = \ket{\psi^+} \otimes \ket{\psi^-}$ 
and $\ket{2_L^2} = \ket{\psi^-} \otimes \ket{\psi^+}$, i.e., we will choose to encode 
$\ket{\Psi_{\text{initial}}} := \cos{\theta} \ket{2_L^2} + \sin{\theta}e^{i\phi} \ket{1_L^2}$. Explicitly, these states take the form
\bea
\label{eq:basis1}
\ket{1_L^2} &:=& (\ket{VHVH}-\ket{VHHV}+\ket{HVVH} \nonumber \\      
&& -\ket{HVHV} )/2 \\
\ket{2_L^2} &:=& (\ket{VHVH}+\ket{VHHV}-\ket{HVVH} \nonumber \\      
&& -\ket{HVHV} )/2
\label{eq:basis2}
\eea

This encoding protects against arbitrary collective noise processes $H_{{\text{error}}} = \sum_{j=0,x,y,z}\sum_{\alpha=1}^4 c_j \sigma_{j}^{(\alpha)}$, where the coefficients $c_j$ govern the relative strength of the $j$th collective Pauli operation $\sum_{\alpha=1}^4\sigma_{j}^{(\alpha)}$ 
(with $\sigma_{0}^{(\alpha)}:=\Bid^{(\alpha)}$). In fact, the state $\ket{\Psi_{DFS}}$ is invariant under the transformation 
$U(\tau)=\exp{(-iH_{{\text{error}}}\tau/\hbar)} = \tilde{U} \otimes \tilde{U} \otimes \tilde{U} \otimes \tilde{U}$ for some unitary 
$\tilde{U}$. The initial coefficients $\nu_i$ remain unchanged as the system evolves under the collective interactions. Although $\nu_i \mapsto \nu_i$, the coefficients $\omega_{i,j}$ generally change, i.e., 
$\ket{{\cal{Q}}_L} = \sum_k \omega_{{\cal{Q}},k}\ket{{\cal{Q}}_L^k} \mapsto \sum_k \omega_{{\cal{Q}},k}^{\prime}\ket{{\cal{Q}}_L^k},
 ({\cal{Q}}=0,1,2)$. The normalization condition $\sum_k |\omega_{{\cal{Q}},k}^{\prime}|^2=1$ is satisfied throughout the 
evolution. 

Although the structure of each logical basis state may change as the system experiences collective noise, each  
logical basis state remains confined to its protected subspace.

\emph{Heralding an arbitrary DFS state.---}
It can be seen that $\ket{1_L^2}$ and $\ket{2_L^2}$ are related through the transformation 
\beq
\ket{1_L^2} =(\sigma_z)_2(\sigma_z)_3\ket{2_L^2},
\eeq
with $\sigma_z \ket{V}=\ket{V}$ and $\sigma_z \ket{H}=-\ket{H}$. Since $\ket{1_L^2}$ and $\ket{2_L^2}$ are tensor products of Bell-states they can each be heralded using two independent heralded Bell-pair sources \cite{Sliwa,Pittman1,Browne1,Joo1,Walther, Zhang,Wagenknecht, Barz}. We will assume that a particular implementation has been arranged to herald the 
logical state $\ket{2_L^2}$. Our objective is to describe an optical circuit which performs the operation 
\beq
O:=\cos{\theta} \Bid + \sin{\theta}e^{i\phi} (\sigma_z)_2(\sigma_z)_3.
\eeq
Our joint phase operation $(\sigma_z)_2(\sigma_z)_3$ 
relies on an extension of the work reported in Ref.~\cite{Pittman}. There, Pittman, Jacobs, and Franson (PJF) 
present probabilistic CNOT and C-Phase gates using polarizing beam splitters. An illustration 
of the PJF C-Phase design is provided in Fig. 1. This setup consists of two polarizing beam splitters and two photon detectors. PBSs sketched with a box and a diagonal line are assumed to transmit $\ket{H}$ and reflect $\ket{V}$. The beam splitters sketched with a box, a diagonal line, and a circle are 
constructed to transmit the polarization state 
$\ket{F} := (\ket{H} +\ket{V})/\sqrt{2}$ and reflect the state $\ket{S}:=(\ket{V} - \ket{H})/\sqrt{2}$. We will refer to these beam splitters as HV-PBSs and FS-PBSs, respectively.
\begin{figure}[!ht]
\label{fig:logic}
\includegraphics[width=.23\textwidth]{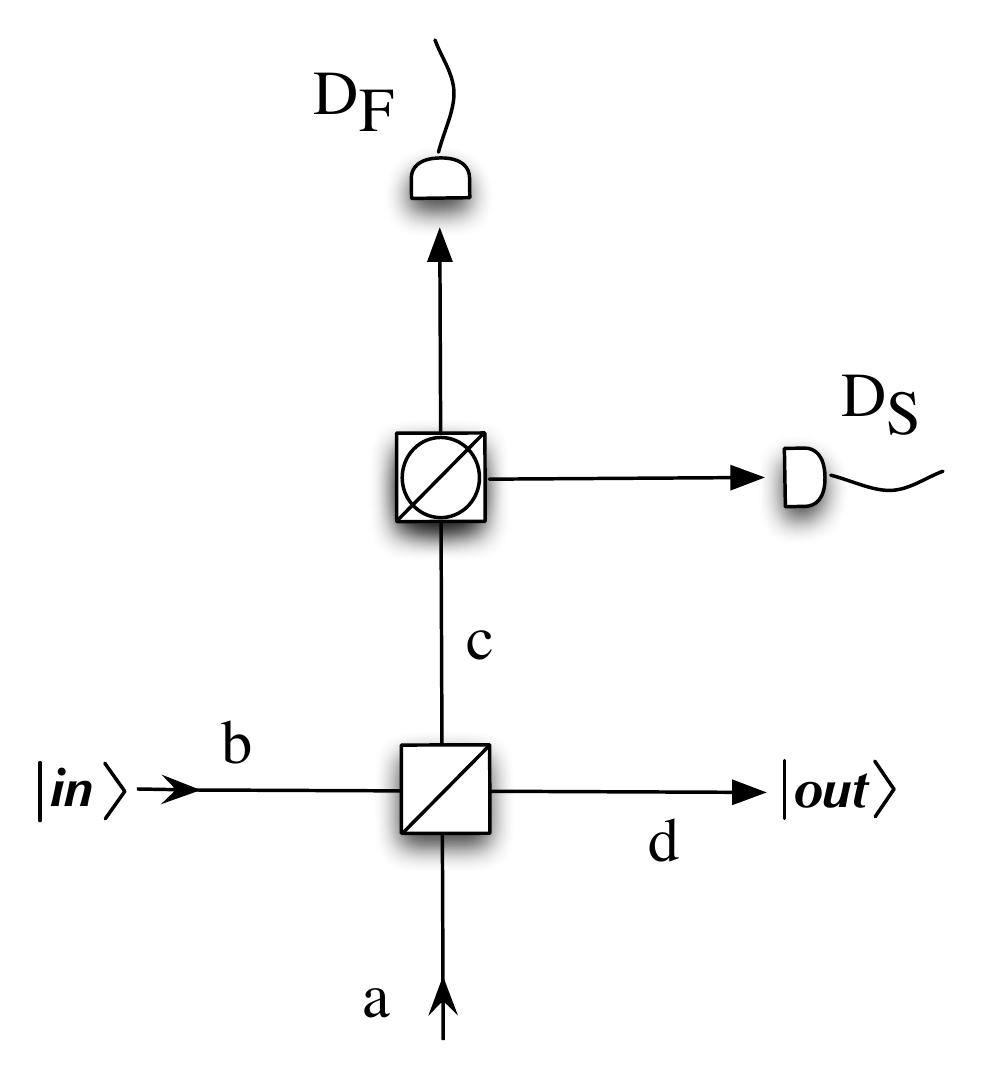}
\caption{Schematic of the Pittman, Jacobs, and Franson quantum parity check 
protocol \cite{Pittman}. For our purposes, we will view this setup as a way to realize probabilistic C-Phase operations.}
\end{figure}
PJF presented this arrangement as a means for performing probabilistic quantum parity check operations. For our purposes, we will view this device as way to implement probabilistic C-Phase 
operations on the target qubit states entering via mode $b$. The target state 
$\ket{in}=\alpha \ket{H_b} +\beta \ket{V_b}$ enters the device along with a second 
photon prepared in the state $\ket{F_a} = (\ket{H_a} +\ket{V_a})/\sqrt{2}$ entering via mode $a$. The combined initial state 
$\ket{\Xi}= (\alpha \ket{H_b} +\beta \ket{V_b}) \otimes (\ket{H_a} +\ket{V_a})/\sqrt{2}$ evolves to 
\bea
\label{eq:PC}
\ket{\Xi} &\rightarrow& \frac{1}{2}\left[\ket{F_{D_F}}(\alpha \ket{H_d}+ \beta \ket{V_d})\right. \nonumber \\
&+& \left. \ket{S_{D_S}}(-\alpha \ket{H_d}+ \beta \ket{V_d})\right] 
+ \frac{1}{\sqrt{2}}\ket{\cal{O_{\text{rej}}}}, 
\eea    
where $\ket{\cal{O_{\text{rej}}}}$ is a normalized state composed of amplitudes which will result in zero or two photons 
being detected. This device therefore allows for the probabilistic application of a $\sigma_z$ operation on the 
target state.  
\begin{figure}[!ht]
\label{fig:NS}
\includegraphics[width=.33\textwidth]{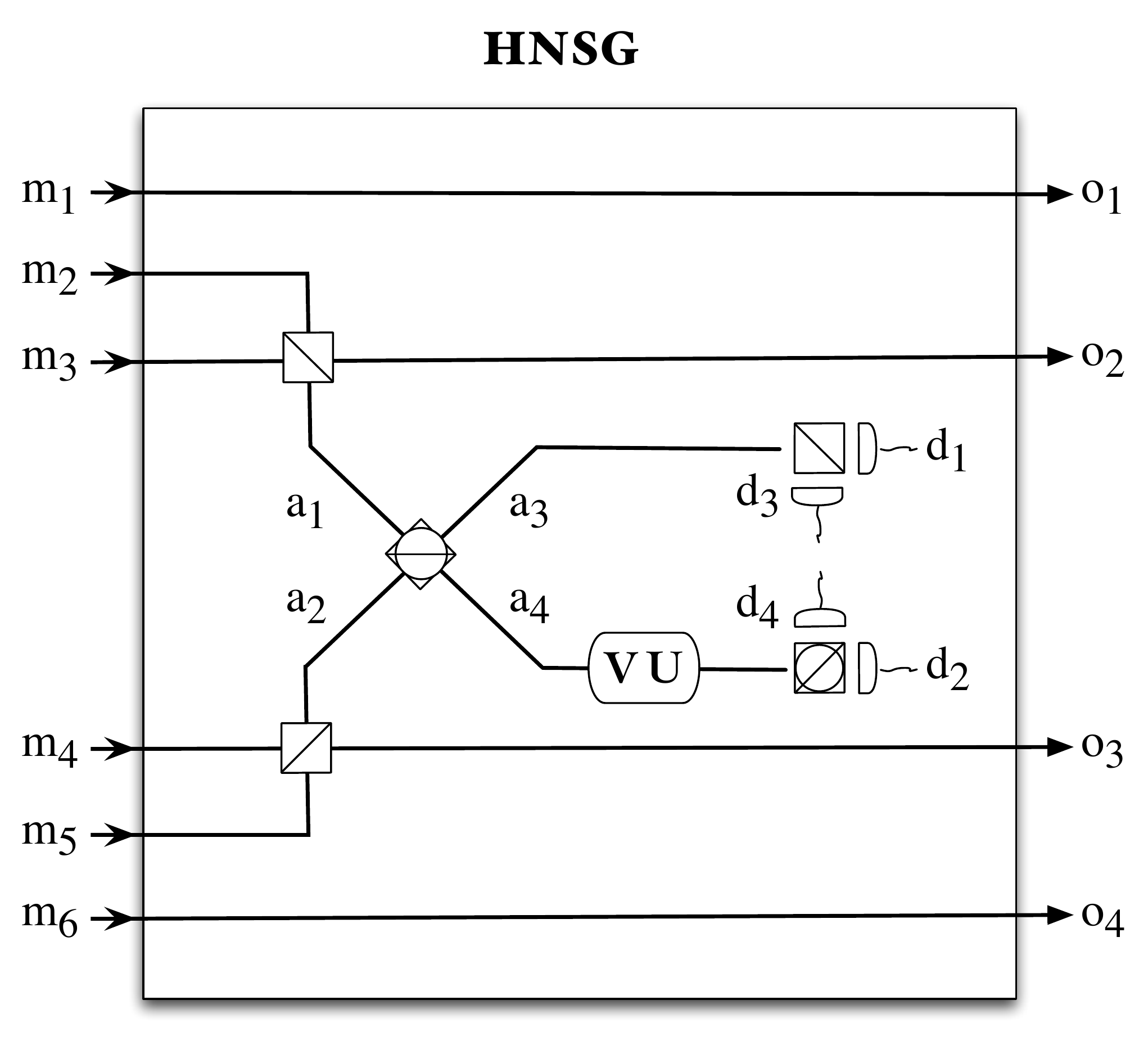} 
\caption{Circuit for generating a heralded and arbitrary DFS qubit state. Successful state preparation is achieved by detecting one and only one photon in modes $d_1$ and $d_2$ while detecting zero photons in modes $d_3$ and $d_4$.}
\end{figure}

We can realize the joint operations $\Bid \otimes \Bid $ and $\sigma_z \otimes \sigma_z $ on two photons by combining two C-Phase gates at a central FS-PBS. This arrangement is incorporated into the heralded noiseless-subsystem generator (HNSG) introduced in 
Fig.~2. In order to clearly explain the details concerning the joint phase operation we will temporarily ignore spatial modes 
$m_1$ and $m_6$ and focus on the evolution of input states originating from modes $m_2,\hdots, m_5$. Suppose two arbitrary 
initial single photon encodings $\ket{in_{m_3}}=\alpha \ket{H_{m_3}}+ \beta \ket{V_{m_3}}$ and $\ket{in^{\prime}_{m_4}}=\alpha^{\prime} \ket{H_{m_4}}+ \beta^{\prime} \ket{V_{m_4}}$ enter the HNSG along with two photons prepared in the states $\ket{F_{m_2}}$ and $\ket{F_{m_5}}$. The combined system evolves to    
\bea
\label{eq:jointphase}
\ket{in_{m_3}}\ket{in^{\prime}_{m_4}}\ket{F_{m_2}}\ket{F_{m_5}} &&\rightarrow \nonumber \\ 
&& \!\!\!\!\!\!\!\!\!\!\!\!\!\!\!\!\!\!\!\!\!\!\!\!\!\!\!\!\!\!\!\!\!\!\!\!\!\!\!\!\!\!\!\!\!\!\!\!\!\!\!\!\!\!\!\!\!\!\!\!\!\!\!\!\!\!\frac{1}{4}\left[\ket{F_{a_3}}\ket{F_{a_4}} \ket{\chi_1} +
 \ket{S_{a_3}}\ket{S_{a_4}} \ket{\chi_2}\right] + \sqrt{\frac{7}{8}}\ket{\cal{O}^{\prime}_{\text{rej}}}, \nonumber \\
\eea
with 
\bea
\ket{\chi_1}&=&(\alpha \ket{H_{o_2}}+ \beta \ket{V_{o_2}})(\alpha^{\prime} \ket{H_{o_3}}+ \beta^{\prime} \ket{V_{o_3}}), \nonumber \\
\ket{\chi_2}&=&(-\alpha \ket{H_{o_2}}+ \beta \ket{V_{o_2}})(-\alpha^{\prime} \ket{H_{o_3}}+ \beta^{\prime} \ket{V_{o_3}}), \nonumber \\
\eea
and $\ket{\cal{O}^{\prime }_{\text{rej}}}$ is a normalized state which does not contain any terms having one and only one (1AO1) photon in modes $a_3$ and $a_4$. We see 
from Eq.~(\ref{eq:jointphase}) that a measurement of 1AO1 photon at each detector with polarization $\ket{F}$ leaves 
the target states unchanged. A measurement of $\ket{S_{a_3}}\ket{S_{a_4}}$ effectively applies 
the joint operation $\ket{out_{o_2}}\otimes\ket{out_{o_3}}=\sigma_z \ket{in_{m_3}} \otimes \sigma_z \ket{in^{\prime}_{m_4}}$. We can selectively produce the logical states $\ket{1_L^2}$ and $\ket{2_L^2}$ using this method. 

In order to produce superpositions of the two logical basis states we transform states occupying modes $a_3$ and $a_4$ according 
to the circuit depicted in Fig.~2. After passing through the central FS-PBS, photon polarization is projected to either $\ket{F}$ or $\ket{S}$. Photons in mode $a_3$ are equally likely to be detected at $d_1$ or $d_3$. This measurement is completely unbiased. The amplitudes of the superposition encoding result from the measurement bias imposed by the polarization rotation 
\beq
U(\theta) = \left(
       \begin{array}{cc}
           \cos{\theta} & -\sin{\theta}  \\

                      \sin{\theta}  & \cos{\theta} \\
       \end{array} \right)
\eeq
which rotates 
\bea
\ket{F} &\mapsto& \cos{\theta}\ket{F} -\sin{\theta}\ket{S} \nonumber \\
\ket{S} &\mapsto& \sin{\theta}\ket{F} +\cos{\theta}\ket{S}.
\eea
This rotation allows one to specify the relative probability of applying either $\Bid \otimes \Bid$ or $\sigma_z \otimes \sigma_z$ 
to the states entering via modes $m_3$ and $m_4$ by deeming the preparation stage successful upon measuring 1AO1 photon in mode $d_2$ and none in $d_4$. In order to apply a relative phase shift associated with these joint operations we first apply the unitary operation 
\beq
V(\phi) = \left(
       \begin{array}{cc}
           1 & 0  \\

                      0  & e^{i\phi} \\
       \end{array} \right)
\eeq
to states in mode $a_4$ before applying $U(\theta)$. The polarization states $\ket{F}$ and $\ket{S}$ are assumed to be ordered such that $V(\phi)\ket{F} =\ket{F}$ and $V(\phi)\ket{s} =e^{i\phi}\ket{S}$. 

As mentioned above, we assume that two heralded Bell-pair generators have been triggered so that the state $\ket{2_L^2} = \ket{\psi^-} \otimes \ket{\psi^+}$ is emitted into the four spatial modes $m_1, m_3, m_4, m_6$, i.e., 
\bea
\ket{{\widetilde{in}}} &=& (\ket{V_{m_1}H_{m_3}V_{m_4}H_{m_6}}+\ket{V_{m_1}H_{m_3}H_{m_4}V_{m_6}} \nonumber \\    
&& \!\!\!\!-\ket{H_{m_1}V_{m_3}V_{m_4}H_{m_6}}   
 -\ket{H_{m_1}V_{m_3}H_{m_4}V_{m_6}} )/2. \nonumber \\ 
\eea
The total input state evolves according to 
\bea
\ket{{\widetilde{in}}}\ket{F_{m_2}}\ket{F_{m_5}} &&\mapsto \nonumber \\
&&\!\!\!\!\!\!\!\!\!\!\!\!\!\!\!\!\!\!\!\!\!\!\!\!\!\!\!\!\!\!\!\!\!\frac{1}{4\sqrt{2}}\left[\ket{F_{d_2}}\ket{V_{d_3}}\left(\cos{\theta}\ket{2_L^2} - e^{i\phi}\sin{\theta}\ket{1_L^2}\right)\right. \nonumber \\
&& \left.\!\!\!\!\!\!\!\!\!\!\!\!\!\!\!\!\!\!
+ \ket{F_{d_2}}\ket{H_{d_1}}\left(\cos{\theta}\ket{2_L^2} + e^{i\phi}\sin{\theta}\ket{1_L^2}\right)\right] \nonumber \\
&& \!\!\!\!\!\!\!\!\!\!\!\!\!\!\!\!\!+\frac{\sqrt{15}}{4}\ket{\cal{O}^{\prime \prime}_{\text{rej}}},
\eea
where $\ket{\cal{O}^{\prime \prime}_{\text{rej}}}$ is a normalized state which is rejected. We can therefore herald a general  
DFS qubit state
\beq
\ket{\Psi_{\text{initial}}} := \cos{\theta} \ket{2_L^2} + \sin{\theta}e^{i\phi} \ket{1_L^2}
\eeq
conditioned on the measurement of the state
\beq
\ket{F_{d_2}}\ket{H_{d_1}}\ket{vac_{d_3}}\ket{vac_{d_4}}.
\eeq

In order to herald the logical state $\ket{0_L}$ we first recognize that
\beq
\ket{0_L^2} = (\sigma_x)_1 (\sigma_x)_4 \left[ \frac{1}{\sqrt{2}}\left(\ket{1_L^2}+\ket{2_L^2}\right)\right].
\eeq
The third DFS qutrit basis state can be heralded by setting $\theta=\pi/4, \phi=0$ and placing wave plates in modes $m_1$ and $m_6$ in order to rotate $\ket{H} \leftrightarrow \ket{V}$. The HNSG therefore has the ability to prepare all three basis states of a DFS qutrit. 
The efficiency of successful state preparation, assuming two Bell-pairs and two unentangled photons each in the state 
$\ket{F}$ have in fact entered the device, is roughly $3.1\%$. This probability should be multiplied by the probability of witnessing two simultaneously heralded Bell-pairs, as well as as two $\ket{F}$ states, in order to obtain the overall preparation efficiency. 
The rate of state generation will be low using current technology since heralded Bell-pair schemes typically produce pairs with 
a low probability of success.

\emph{Decoding the logical states.---}
Although there is a great deal of freedom in the DFS initialization process, a decoding mechanism must have the ability to distinguish every form of a given basis state from the other logical basis states.
\begin{figure}[!ht]
\label{fig:NS}
\includegraphics[width=.3\textwidth]{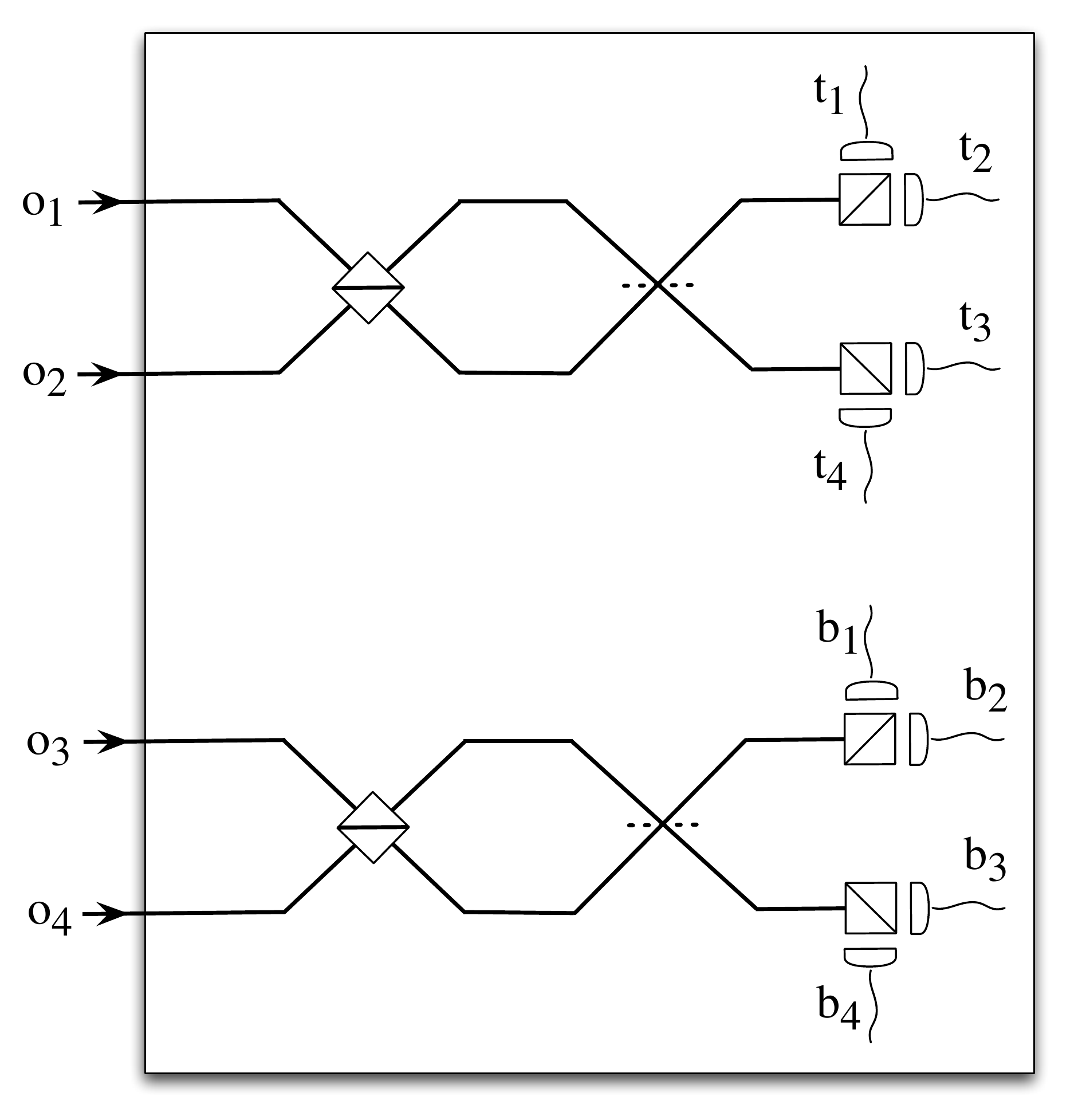} 
\caption{Schematic of an interferometer which can decode the logical basis states of a four-qubit noiseless subsystem. The setup consists of several HV-PBSs, two ordinary 50/50 beam-splitters, and eight photon detectors.}
\end{figure}
In other words, we must take into account {\it{all}} 
possible collective-noise channels $\ket{{\cal{Q}}_L} = \sum_k \omega_{{\cal{Q}},k}\ket{{\cal{Q}}_L^k} \mapsto \sum_k \omega_{{\cal{Q}},k}^{\prime}\ket{{\cal{Q}}_L^k},
 ({\cal{Q}}=0,1,2)$. In general, a receiver can expect to receive 
\bea
\ket{1_L}&=& \left(\alpha_1 \ket{VV}+ \beta_1 \ket{\psi^+}+\gamma_1\ket{HH}\right)\otimes \ket{\psi^-} \nonumber \\
\ket{2_L}&=& \ket{\psi^-} \otimes \left(\alpha_2 \ket{VV}+ \beta_2 \ket{\psi^+}+\gamma_2\ket{HH}\right). \nonumber \\
\eea

Fortunately, these states are separable. A decoder which can distinguish $\ket{\psi^-}$ from the set 
$\{\ket{VV},\ket{HH}, \ket{\psi^+}\}$ will suffice for decoding in one basis. The interferometer depicted in Fig.~3 has this ability. 
This setup consists of two identical parts, one for modes $o_1$ and $o_2$, and the other for $o_3$ and $o_4$. Since these parts are identical, we will only focus on one of them. Consider the top portion consisting of two input modes $o_1$ and $o_2$, an HV-PBS, an ordinary 50/50 beam-splitter, two additional HV-PBSs, and four detectors. It can be shown that the input state $\ket{\psi^-_{o1,o2}}$ leads to detector clicks at either $(t_1,t_2)$ or $(t_3,t_4)$. The 
states within the set $\{\ket{VV},\ket{HH}, \ket{\psi^+}\}$ can be shown to yield the following detection events: $\ket{V_{o_1}V_{o_2}} \Rightarrow (t_1,t_1)$ or $(t_4,t_4)$, $\ket{H_{o_1}H_{o_2}} \Rightarrow (t_2,t_2)$ or $(t_3,t_3)$, and $\ket{\psi^+_{o1,o2}} \Rightarrow (t_2,t_4)$ or $(t_1,t_3)$. Here, $(t_1,t_1)$ means that two photons are detected at $t_1$,  $(t_3,t_4)$ means that one photon is detected at $t_3$ and another at $t_4$, etc. Identical results hold for the bottom portion. This setup can easily distinguish the states $\ket{1_L}$ and 
$\ket{2_L}$ since these measurement outcomes are distinct. Furthermore, it can decode $\ket{0_L}$ as well given the fact that the 
space which spans $\ket{0_L}$ does not contain a $\ket{\psi^-}$ contribution.

\emph{Conclusions.---}
We have presented a proposal for an optical device that is capable of heralding an arbitrarily encoded decoherence-free qubit. Our device takes as input two heralded Bell-pairs, as well as two unentangled photons, and outputs the appropriate state conditioned on the detection of two auxiliary photons. Arbitrary state preparation is achieved using a single polarization rotator and a single birefringent phase shifter along with four number-resolving photon detectors. Alternatively, the setup can also be used to postselect an arbitrary DFS qubit state with a higher efficiency compared to the heralding case. For postselection, modes $m_1,m_3,m_4$ and $m_6$ are matched to the signal and idler modes of 
two down-conversion sources. Successful postselection results from the simultaneous detection of one and only one photon in modes $o_1,o_2,o_3$ and $o_4$, along with the appropriate detector clicks which accompany the heralding scheme. The setup can tolerate unwanted multiple down-conversions in a single crystal since modes $m_1$ and $m_6$ never overlap with any other photon paths.

An interferometric decoding device which can distinguish all three DFS  basis states in the logical basis was also provided. This allows one to determine the amplitudes associated with an arbitrary superposition encoding. The problem of decoding the logical qubit state in three mutually unbiased bases remains an open question. Finding a proper configuration which will allow for the preparation of an arbitrarily encoded DFS qutrit state remains to be seen as well.

\emph{Acknowledgements.---}
This work was performed at Oak Ridge National Laboratory, operated by UT-Battelle for the U.S. Department of Energy. The work has been authored by a contractor of the U.S. Government. Accordingly, the U.S. Government retains a nonexclusive, royalty-free license to publish or reproduce the published form of this work, or to allow others to do so for U.S. Government purposes.



\end{document}